\documentclass[twocolumn,showpacs,superscriptaddress,prd,amsmath,amssymb,floatfix]{revtex4-1}
\usepackage{graphicx}
\usepackage{epsfig}
\usepackage{dcolumn}
\usepackage{bm}
\usepackage{overpic}
\usepackage{color}
\usepackage{multirow}
\usepackage{subfigure}
\newcommand{\jpsi}{J/\psi}

\newcommand{\bfg}{\begin{figure}}
\newcommand{\efg}{\end{figure}}
\newcommand{\bitm}{\begin{itemize}}
\newcommand{\eitm}{\end{itemize}}
\newcommand{\bnum}{\begin{enumerate}}
\newcommand{\enum}{\end{enumerate}}
\newcommand{\btbl}{\begin{table}}
\newcommand{\etbl}{\end{table}}
\newcommand{\btbu}{\begin{tabular}}
\newcommand{\etbu}{\end{tabular}}
\newcommand{\bcl}{\begin{center}}
\newcommand{\ecl}{\end{center}}

\newcommand{\beq}{\begin{equation}}
\newcommand{\eeq}{\end{equation}}
\newcommand{\beqr}{\begin{eqnarray}}
\newcommand{\eeqr}{\end{eqnarray}}

\def \jpsi    {J/\psi}
\def \psip    {\psi(3686)}
\def \epem    {e^+e^-}
\def \D0      {D^{0}}
\def \piz     {\pi^0}

\def \kpi     {K^-\pi^+}
\def \kpipi   {K^-\pi^+\pi^0}
\def \kpipipi {K^-\pi^+\pi^+\pi^-}
\def \pppm    {\pi^+\pi^-}
\def \Br   {\mathcal{B}}

\def \gevcc{\mbox{GeV/$c^2$}}
\def \mev  {\mbox{MeV}}
\def \mevcc{\mbox{MeV/$c^2$}}

\begin{document}
\title{\boldmath Search for the rare decays $\jpsi \to \D0 \epem +c.c.$ and $\psip \to \D0 \epem +c.c.$}
\author{
\begin{small}
M.~Ablikim$^{1}$, M.~N.~Achasov$^{9,d}$, S.~Ahmed$^{14}$, M.~Albrecht$^{4}$, A.~Amoroso$^{50A,50C}$, F.~F.~An$^{1}$, Q.~An$^{47,39}$, J.~Z.~Bai$^{1}$, O.~Bakina$^{24}$, R.~Baldini Ferroli$^{20A}$
% ...
\\
\vspace{0.2cm}
(BESIII Collaboration)\\
\vspace{0.2cm} {\it
$^{1}$ Institute of High Energy Physics, Beijing 100049, People's Republic of China\\
% ...
$^{4}$ Bochum Ruhr-University, D-44780 Bochum, Germany\\
% ...
$^{9}$ G.I. Budker Institute of Nuclear Physics SB RAS (BINP), Novosibirsk 630090, Russia\\
% ...
$^{14}$ Helmholtz Institute Mainz, Johann-Joachim-Becher-Weg 45, D-55099 Mainz, Germany\\
% ...
$^{20}$ (A)INFN Laboratori Nazionali di Frascati, I-00044, Frascati, Italy; (B)INFN and University of Perugia, I-06100, Perugia, Italy\\
% ...
$^{24}$ Joint Institute for Nuclear Research, 141980 Dubna, Moscow region, Russia\\
% ...
$^{39}$ State Key Laboratory of Particle Detection and Electronics, Beijing 100049, Hefei 230026, People's Republic of China\\
% ...
$^{47}$ University of Science and Technology of China, Hefei 230026, People's Republic of China\\
% ...
$^{50}$ (A)University of Turin, I-10125, Turin, Italy; (B)University of Eastern Piedmont, I-15121, Alessandria, Italy; (C)INFN, I-10125, Turin, Italy\\
% ...
$^{d}$ Also at the Novosibirsk State University, Novosibirsk, 630090, Russia\\
% ...
 }
\vspace{0.4cm}
}
\noaffiliation{}

\date{\today}

\begin{abstract}

  Using the data samples of $(1310.6\pm7.2 )\times 10^{6}$ $\jpsi$ events and $(448.1\pm2.9)\times 10^{6}$ $\psip$ events collected with the BESIII detector, we search for the rare decays $\jpsi\to\D0 \epem + c.c.$ and $\psip\to\D0 \epem + c.c.$.
  No significant signals are observed and the corresponding upper limits on the branching fractions at the $90\%$ confidence level are determined to be $\Br(\jpsi \to \D0 \epem + c.c.)< 8.5\times 10^{-8}$ and $\Br (\psip \to \D0 \epem + c.c.)<1.4\times 10^{-7}$, respectively.
  Our limit on $\Br(\jpsi\to \D0 \epem + c.c.)$ is more stringent by
  two orders of magnitude than the previous results, and the $\Br(\psip\to\D0 \epem + c.c.)$ is measured for the first time.
\end{abstract}

\maketitle

In the Standard Model (SM), decays of the charmonium resonances $\jpsi$ and $\psip$ (collectively referred to as $\psi$ throughout the text) induced by Flavor Changing Neutral Current (FCNC) are forbidden at the tree level due to the Glashow-Iliopoulos-Maiani (GIM) mechanism~\cite{GIM}, but can occur via a $c \to u$ transition at the loop level, \emph{e.g.}, shown in Fig.~\ref{feynman} for the decay of $\psi~\to \D0 \epem$. Such decays can also occur via long-distance effects on the hadron level, which are, according to Ref.~\cite{FCNC}, expected to have the same order of magnitude as the FCNC process.
  \begin{figure}[htb]
  \vskip  -0.3cm
  \hskip -0.0cm
  \begin{overpic}[width=0.32\textwidth, height=0.15\textwidth]{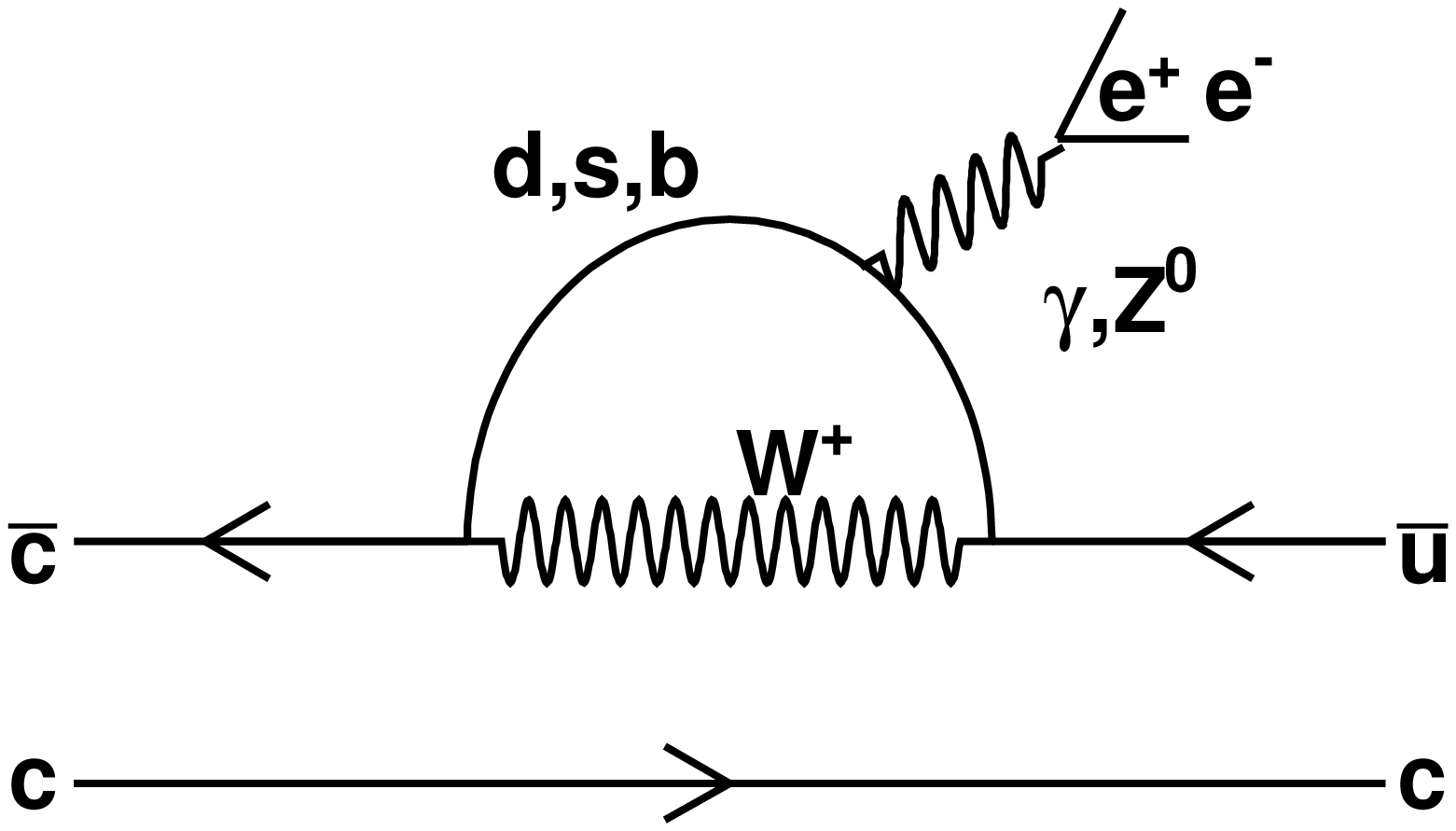}
  \end{overpic}
  \caption{Typical Feynman diagram for $\psi \to D^{0} \epem$.}\label{feynman}
  \end{figure}
The decay branching fraction for this kind of rare process is expected to be of order $10^{-10}$ to $10^{-13}$ in the SM \cite{SM1,SM2}.
However, many new physics models, such as the Topcolor models~\cite{TopC}, the minimal supersymmetric standard model with R-parity violation~\cite{Supersy}, and the two Higgs doublet model~\cite{Thiggs}, predict that the decay branching fractions can be enhanced by two or three orders of magnitude.
Searching for experimental evidence for these FCNC processes offers an ideal opportunity to study non-perturbative QCD effects and their underlying dynamics, and serves as a probe to search for new physics beyond the SM~\cite{newphy1,newphy2}.

The semileptonic decay $J/\psi$ $\to \D0 \epem$ (the charged conjugate channel is always implied throughout the text if not mentioned explicitly) is an interesting decay to study FCNC-induced processes.
The BESII experiment has searched for the decay $\jpsi \to \D0 \epem$ by using $58 \times 10^6$ $\jpsi$ events, and an upper bound on the branching fraction was placed at the order of $10^{-5}$~\cite{besII}, which is far away from the theoretical prediction.
A more precise measurement with larger statistics is desirable to test the theoretical predictions more stringently.

The Beijing Spectrometer (BESIII) detector~\cite{detector1}, located at the Beijing Electron-Positron Collider (BEPCII), has collected ($1310.6 \pm 7.2$) $\times$ $10^{6}$ $\jpsi$ events~\cite{jpsinum1,jpsinum2} and ($448.1 \pm 2.9$) $\times$ $10^{6}$ $\psip$ events~\cite{psipnum1,psipnum2}, which are the world largest samples collected with electron-positron collision at threshold.
The high quality and large statistics data samples provide a unique opportunity to search for physics beyond the SM.
In this paper, we present a search for the rare processes $\psi\to \D0 \epem$. 

BEPCII is a double-ring $\epem$ collider running at center-of-mass (c.m.)~energies between 2.0 and 4.6 GeV,
reaching a peak luminosity of $1.0\times10^{33}$~cm$^{-2}$s$^{-1}$ at a c.m.~energy of 3770 MeV.
The cylindrical BESIII detector has an effective geometrical acceptance of $93\%$ of 4$\pi$ and is divided into a barrel section and two endcaps.
It contains a small-cell, helium-based (40$\%$ He, 60$\%$ C$_{3}$H$_{8}$) main drift chamber (MDC)
which provides momentum measurements for charged particles with a resolution of $0.5\%$ at a momentum
of 1 GeV/$c$ in a magnetic field of 1 Tesla.
The energy loss measurement ($dE/dx$) provided by the MDC has a resolution better than $6\%$.
A time-of-flight system (TOF) consisting of 5-cm-thick plastic scintillators can measure the flight
time of charged particles with a time resolution of 80 ps in the barrel and 110 ps in the endcaps.
An electromagnetic calorimeter (EMC) consisting of 6240 CsI(Tl) crystals in a cylindrical structure and two
endcaps is used to measure the energies of photons and electrons.
The energy resolution of the EMC is $2.5\%$ in the barrel and $5.0\%$ in the endcaps for
photons and electrons with an energy of 1~GeV.
The position resolution of the EMC is 6 mm in the barrel and 9 mm in the endcaps. A muon counter (MUC) system
consisting of more than 1200~m$^{2}$ of Resistive Plate Chambers (RPC) is used to identify muons and provides
a spatial resolution better than 2~cm. A detailed description of the BESIII detector can be found in Ref.~\cite{detector1}.

Large samples of Monte Carlo (MC) simulated events are used to optimize the event selection criteria, estimate the background contaminations and determine the selection efficiencies. The MC samples are generated using a {\sc geant4}-based~\cite{geant4} simulation software package {\sc BESIII object oriented simulation tool (BOOST})~\cite{Deng}, which includes the description of geometry and material, the detector response and the digitization model, as well as tracking for the detector running conditions and performances.
'Inclusive' MC samples of $1225\times 10^{6}$ $\jpsi$ and $506\times 10^{6}$ $\psip$ generic decay events are generated to study the backgrounds.
In the simulation, production of the $\psi$ resonances is simulated with the {\sc kkmc}~\cite{kkmc1} generator, while the decays are generated by {\sc evtgen}~\cite{evtgen} for the known decay modes, setting the branching fractions according to the Particle Data Group (PDG)~\cite{PDG}, or {\sc lundcharm}~\cite{lundcharm} for the remaining unknown decays.
The QED Final State Radiation (FSR) effect is simulated with {\sc photos}~\cite{photos}.
Signal samples $\psi \to \D0 \epem$ of $2.0 \times 10^{5}$ events are generated according to a theoretical calculation based on the vector meson dominance (VMD) model~\cite{DIYg}, where $\epem$ is generated from a virtual photon decay.

To study the decay $\psi\to\D0 \epem$, we reconstruct the $\D0 $ signal through its three prominent exclusive hadronic decay modes, $\kpi$ (mode I), $\kpipi$ (mode II), and $\kpipipi$ (mode III), which have relatively large branching fractions, and suffer from relatively low background.

Charged tracks are reconstructed from hits in the MDC. A good charged track is required to have a polar angle $\theta$ that satisfies $|\cos\theta| < 0.93$ and a distance of closest approach point to the interaction point (IP) within 10~cm along the beam direction and 1~cm in the plane perpendicular to the beam.
The measured ionizing energy loss $dE/dx$ in the MDC and flight time in the TOF are combined to form a particle identification (PID) confidence level (C.L.) for each particle hypothesis $i$ ($i$ = $e$, $\pi$, $K$, $p$).
Each track is assigned to the particle type with the highest C.L.

Photons are reconstructed from clusters of energy depositions in the EMC crystals.
The energy deposited in the nearby TOF counters is included to improve the reconstruction efficiency and energy resolution.
Good photons are required to have energy greater than 25~$\mev$ in the barrel region ($|\cos\theta| < 0.80$) or 50~$\mev$ in the endcap region ($0.86 < |\cos\theta| < 0.92$).
The showers in the transition region between the barrel and endcap are poorly reconstructed and are excluded from the analysis.
To exclude showers from charged particles, a photon must be separated from any charged track by more than $10^{\circ}$.
A requirement on the EMC time $t$ with respect to the event start time of $0\leq \rm{t} \leq 700$~ns is used to suppress electronic noise and energy deposits in the EMC unrelated to the events.

The candidate events are selected by requiring an electron-positron pair, a kaon, one or three pions depending on the $\D0 $ decay mode as well as two photons for decay mode II.
To suppress backgrounds, the electron or positron are required to satisfy $E/p>0.8$, where $E$ and $p$ are the corresponding energy deposited in the EMC and momentum measured in the MDC, respectively.
A vertex fit is performed on the selected charged tracks to ensure that the events originate at the IP.
To improve resolution and reduce backgrounds, a four-constraint (4C) kinematic fit imposing energy-momentum conservation is carried out under the hypothesis of $\psi\to \kpi \epem$ (mode I) or $\psi\to\kpipipi\epem$ (mode III), and $\chi^2_{4C}<60$ is required.
For decay mode II, a 5C kinematic fit is performed under the hypothesis of $\psi\to \kpi \gamma\gamma \epem$ with an additional constraint on the mass of the $\gamma\gamma$ pair to the $\pi^0$ nominal mass (5C), and $\chi^2_{5C}<70$ is required.
For events with more than two photon candidates, the $\gamma\gamma$ combination with the least $\chi^2_{5C}$ is retained for further analysis.

With the above selection criteria, the dominant backgrounds are from processes with similar hadronic final states as the signal but with an $\epem$ pair which comes from a $\gamma$ conversion from interactions with the detector material. This effect occurs primarily at the beam pipe, which has an inner diameter of 63~mm, or the inner MDC wall, which has a diameter of 108~mm. To suppress these backgrounds, a $\gamma$ conversion finder algorithm~\cite{xuzr} was developed to reconstruct the $\gamma$ conversion vertex. A set of parameters is defined to study the properties of the $\gamma$ conversion vertex. These include the distance $R_{xy}$ from the IP to the reconstructed vertex point of the $\epem$ pair in the $\it{x}$-$\it{y}$ plane, the distance $\Delta_{\it{xy}}$ between the two intersection points of the two circles describing the trajectories of the electron and positron in the $\it{x}$-$\it{y}$ plane and the line connecting their centers,
the invariant mass $M(\epem)$ of the $\epem$ pair,
the angle $\theta_{eg}$ between the photon momentum vector (in the $\epem$ system) and the direction from the IP (run averaged) to the reconstructed vertex in the $\it{x}$-$\it{y}$ plane, and the angle $\theta_{\epem}$ between the momenta of $e^{+}$ and $e^{-}$ in the $\it{x}$-$\it{y}$ plane.
The $\gamma$ conversion vertex is identified with the criteria $R_{\it{xy}}>2$~cm and $\Delta_{\it{xy}}<6\sigma_{\Delta_{\it{xy}}}$ as well as $\cos\theta_{eg}>0.99$, and the events satisfying these criteria are removed.  This selection removes 95.0\% of all $\gamma$ conversion events while losing less than 5.0\% efficiency for the signal decays.  The criteria are determined by studying a control sample of $\jpsi\to\gamma \pppm$ with the $\gamma$ conversion into a $\epem$ pair.  The resolution $\sigma_{\Delta_{\it{xy}}}$  of $\Delta_{\it{xy}}$ depends on the angle $\theta_{\epem}$ and is also determined from the study of the control sample.

After the $\gamma$ conversion suppression criteria are applied, the inclusive $\jpsi$ and $\psip$ MC samples are used to study the remaining background contamination.
In decay mode I, the dominant backgrounds are $\psi\to (\gamma) \pppm\epem$ $\it{etc}$., due to $K$/$\pi$ mis-identification for high momentum tracks.
The 4C kinematic fits with hypotheses $\psi\to(\gamma)\pppm\epem$ are performed, and the corresponding $\chi^{2}_{4C}(\kpi\epem) < \chi^{2}_{4C}((\gamma)\pppm\epem)$ is required.
Another potential background in the $\psip$ data sample is $\psip\to\pppm\jpsi$ with subsequent decay $\jpsi\to\epem$, where a low momentum $\pi^{\pm}$ is mis-identified to be an electron while a high momentum electron (positron) is mis-identified as a kaon.
The requirement $|M^{\rm Recoil}(\pppm)-m(\jpsi)|<0.02~\gevcc$ is implemented to reduce this background, where $M^{\rm Recoil}(\pppm)$ is the recoil mass of the two low momentum tracks with opposite charges using the $\pi^{\pm}$ hypothesis,
and $m(\jpsi)$ is the nominal $\jpsi$ mass.
In decay mode II, the dominant background is $\psi \to K_{S} \kpi$ with subsequent decays $K_{S}\to\pi^0\pi^0$ and with a $\pi^0$ Dalitz decay.
The requirement $|M^{\rm Recoil}(\kpi)-m(K_S)|>0.06~\gevcc$ is applied to suppress this background, where $M^{\rm Recoil}(\kpi)$ is the recoil mass of the $\kpi$ system, and $m(K_S)$ is the $K_S$ nominal mass.
Another background is $\psi\to\omega \pppm$ ($\omega\to\pi^0\epem$) with a pion mis-identified as a kaon. This background is suppressed by a requirement on the kinematic fit quality with $\chi^{2}_{5C}(\kpi\piz\epem) < \chi^{2}_{5C}(\pppm\piz\epem)$.
In decay mode III, the dominant background is $\psi\to\epem K_{S} K\pi$ with the subsequent $K_S$ decay into $\pppm$. This background is rejected if any of the $\pppm$ invariant mass $M(\pppm)$ satisfies $|M(\pppm)-m(K_{S})|<0.025~\gevcc$.

\begin{figure}

\setlength{\abovecaptionskip}{0.2cm}
\setlength{\belowcaptionskip}{-0.5cm}
 \centering
 \begin{overpic}[width=0.23\textwidth, height=0.20\textwidth]{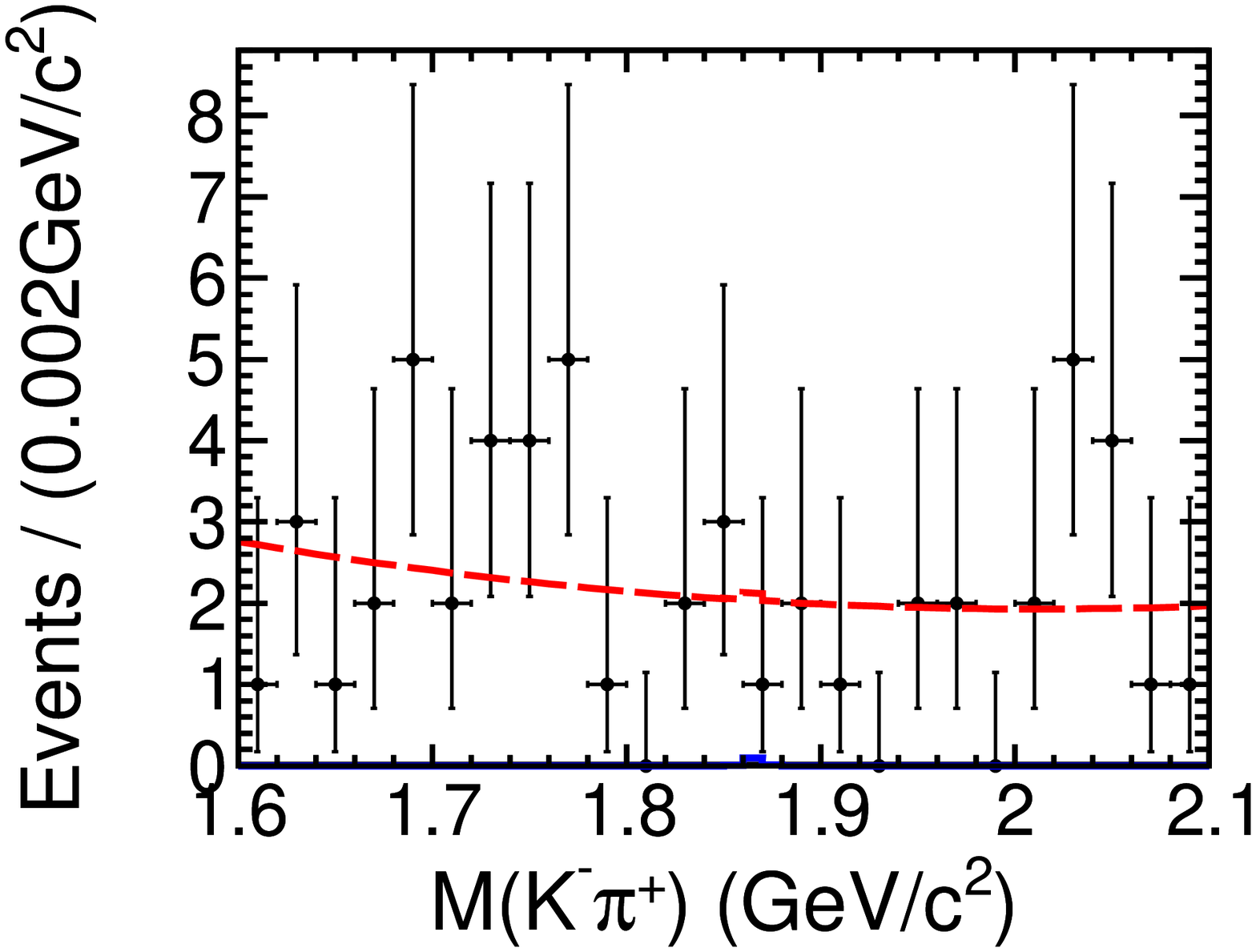}
 \end{overpic}
 \begin{overpic}[width=0.23\textwidth, height=0.20\textwidth]{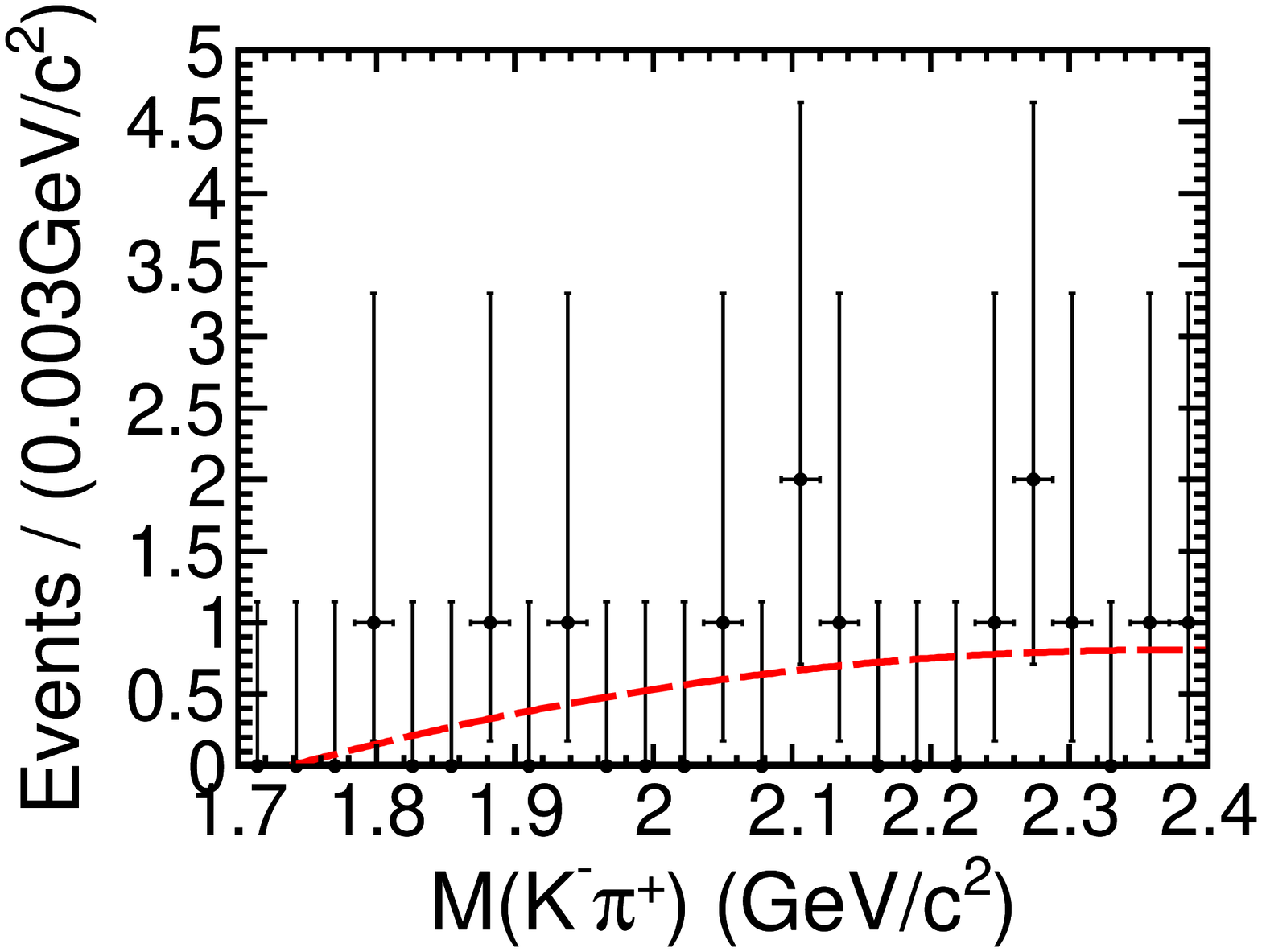}
 \end{overpic}
  \begin{overpic}[width=0.23\textwidth, height=0.20\textwidth]{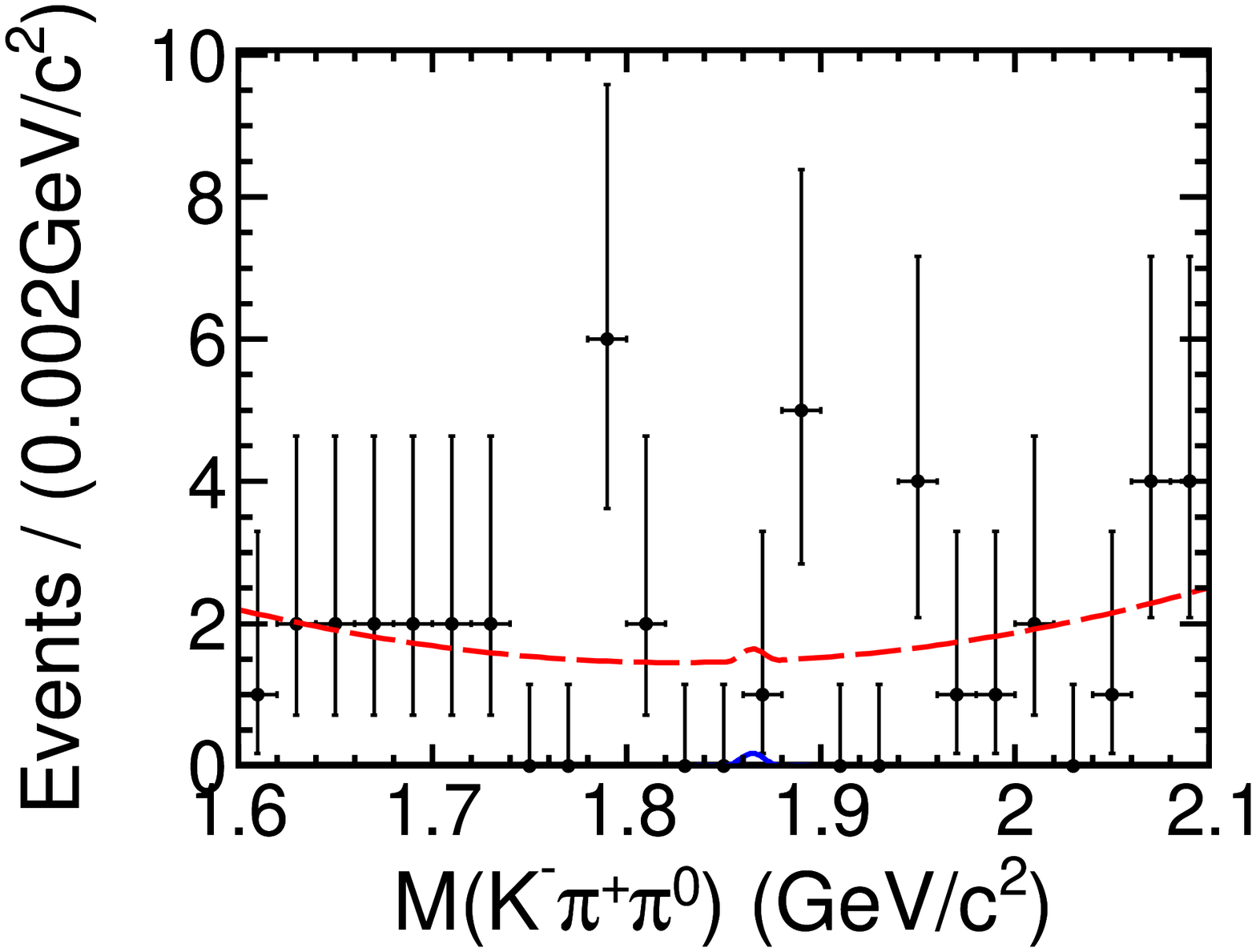}
 \end{overpic}
 \begin{overpic}[width=0.23\textwidth, height=0.20\textwidth]{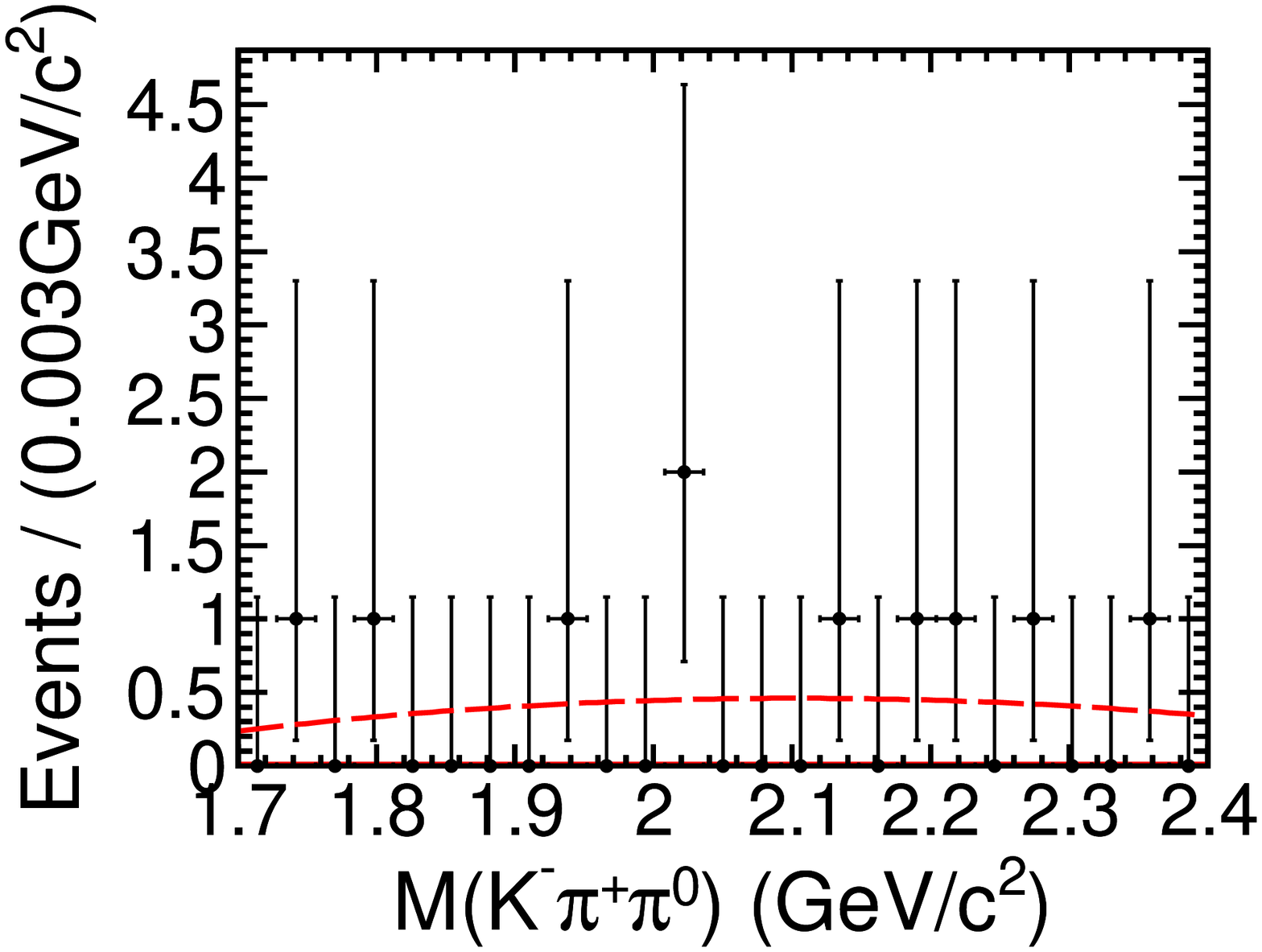}
 \end{overpic}
  \begin{overpic}[width=0.23\textwidth, height=0.20\textwidth]{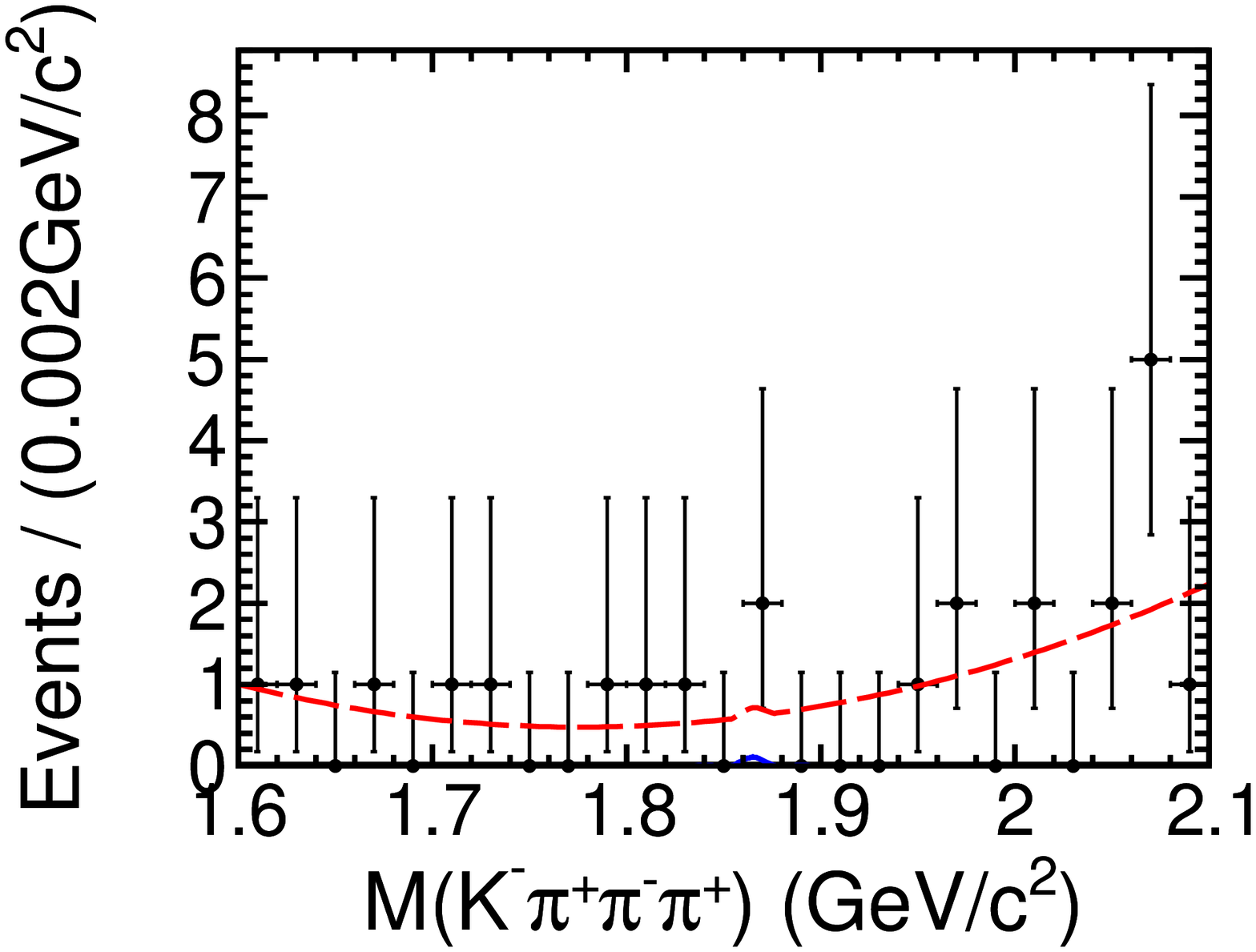}
 \end{overpic}
 \begin{overpic}[width=0.23\textwidth, height=0.20\textwidth]{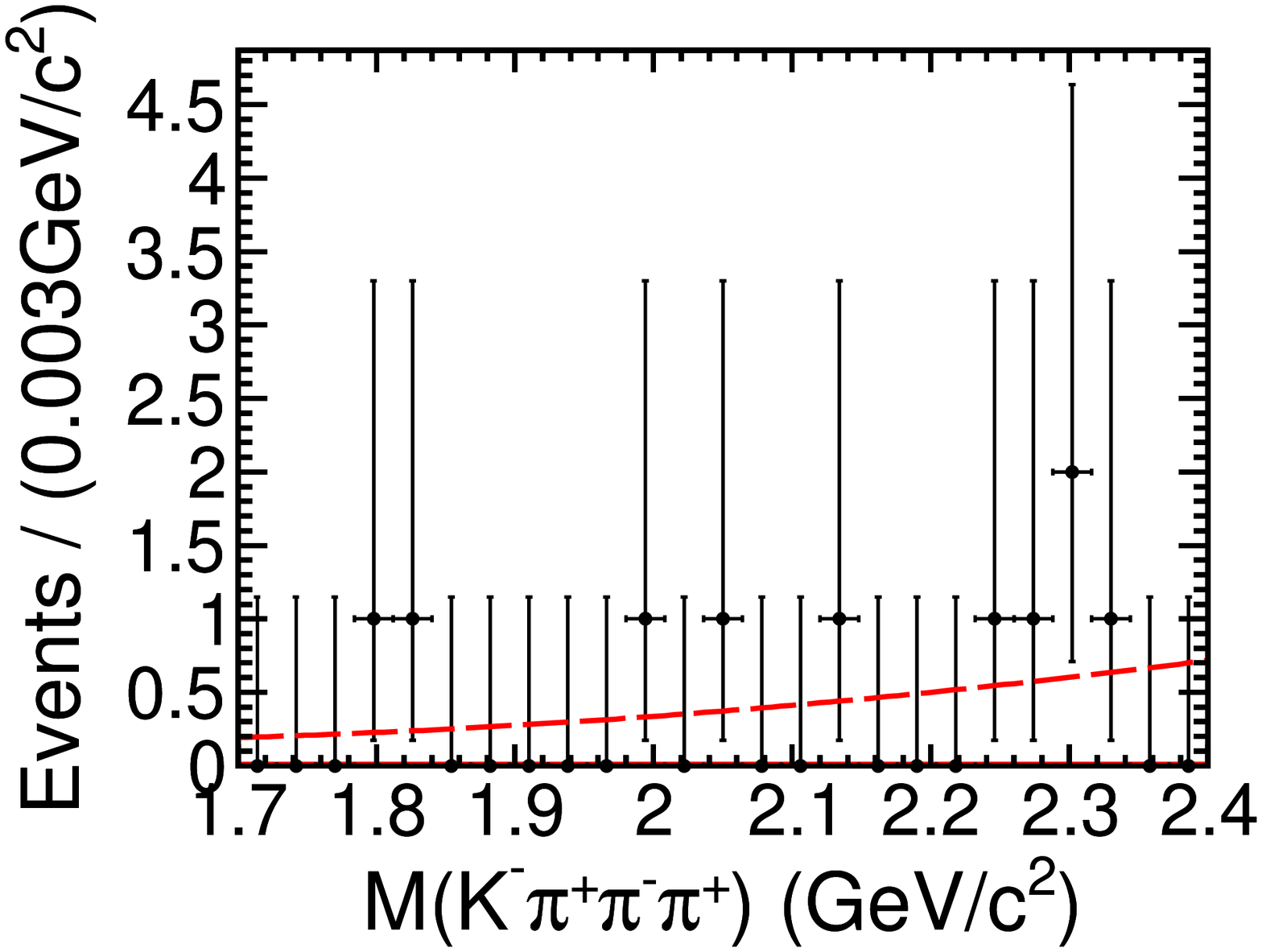}
 \end{overpic}
  \caption{ Distributions of $\kpi$ (upper row), $\kpi\piz$ (middle row) and $\kpi\pppm$ (bottom row) invariant masses.
  The left and right columns are for the $\jpsi$ and $\psip$ samples, respectively.
  Dots with error bars are data, the solid and dashed curves are for the signal shape and the total best fit to data, respectively.}
 \label{masspsi}
\end{figure}

After applying the above selection criteria, the distributions of the $\kpi$, $\kpipi$, and $\kpipipi$ invariant masses for the surviving events in the three $\D0 $ meson decay modes are shown in Fig.~\ref{masspsi}.
No $\D0 $ signals are observed, and therefore upper limits on the branching fractions at the 90\% C.L. are determined.

In the measurements of the branching fractions, the sources of systematic uncertainty include the detection efficiencies of charged tracks and photons, the PID efficiency, the kinematic fit, $\gamma$ conversion veto, mass window requirements, the fit procedure, the decay branching fractions of intermediate states, as well as the total numbers of $\psi$ events.
The individual systematic uncertainties are estimated and described in detail as follows:

{\it (a) Tracking efficiency:}
The tracking efficiencies for $\pi^\pm$ and $K^\pm$ are studied using control samples of $\jpsi \to \rho \pi \to \pppm \piz$,
$\jpsi \to p\bar{p}\pppm$, and $\jpsi \to K^{0}_{S}\kpi$~\cite{trackerror1,trackerror2}.
The tracking efficiency for electrons (positrons) is studied with a control sample of radiative Bhabha events.
The differences in tracking efficiencies between data and MC simulation are 1\% per track for $K$, $\pi$ and $e$, respectively, and are taken as the systematic uncertainties.

{\it (b) PID:}
The PID efficiencies of $\pi^\pm$ and $K^\pm$ are studied with the same control samples as in the study of the tracking efficiency~\cite{trackerror1,trackerror2}.
The PID efficiency from the data sample agrees with that of the MC simulation within 1\% for each track.
The uncertainty of the PID efficiency for electrons (positrons) is studied with the control sample of radiative Bhabha events, and 1.0\% is assigned for each electron (positron).
The uncertainty of the $E/p$ requirement for electrons (positrons) is studied with the control sample $\jpsi\to\pppm\piz$ ($\piz\to\gamma\epem$), and an uncertainty of 2\% is assigned.

{\it (c) Photon detection efficiency:}
The photon detection efficiency is studied with the control samples $\jpsi\to\pi^{+}\pi^{-}\piz$, and a weighted average uncertainty, according to the energy distribution, is determined to be 0.6\% per photon.

{\it (d) Kinematic fit:}
The uncertainty associated with the kinematic fit arises from the inconsistency of the track helix parameters between data and MC simulation.
Therefore, the three track parameters $\phi_0$, $\kappa$ and $\tan\lambda$ are corrected for the signal MC samples, where the correction factors are obtained by comparing the pull distributions of the control samples described in detail in Ref.~\cite{kinematic}.
The resulting difference in the detection efficiencies between the samples with and without the helix correction is taken as the systematic uncertainty.

{\it (e) $\gamma$ conversion veto:}
The effect of the $\gamma$ conversion veto is studied using a control sample of $\jpsi \to \pppm \piz$ with the subsequent Dalitz decay $\piz\to \gamma \epem$.
A clean control sample is selected, and the corresponding MC sample is generated with the RhoPi generator based on a formalism of helicity coupling amplitudes for the process $\jpsi\to \pppm\piz$~\cite{rhopi},
while a generator for the decay $\piz\to\gamma\epem$ adopts a simple pole approximation in the form factor $|F(q^{2})| =1+\alpha q^{2}/m^{2}_{\pi^{0}}$ with $\alpha=0.032$~\cite{PDG}.
The efficiency of the $\gamma$ conversion veto is the ratio of signal yields with and without the $\gamma$ conversion veto, where the signal yields are extracted by fitting the $\epem$ invariant mass.
The resulting difference between data and MC, 1.7\%, is taken as the systematic uncertainty.

{\it (f) Mass window requirements:}
Various requirements of mass window by widening 5 $\mevcc$ are applied to veto the different backgrounds, the corresponding uncertainties are studied by changing the appropriate values. The resulting changes in the final results are taken as the systematic uncertainties.

{\it (g) Branching fractions of intermediate states: }
The uncertainties of the decay branching fractions of intermediate states in the cascade
decays are quoted from the PDG~\cite{PDG}.

{\it (h) Total number of $\psi$ events:}
The uncertainties on the total numbers of $\jpsi$ and $\psip$ events are 0.55\% and 0.62\%, respectively, which are determined by studying the inclusive hadron events~\cite{jpsinum1, jpsinum2, psipnum1, psipnum2}.

All the individual systematic uncertainties are summarized in Table~\ref{tab:uncertaintypsi}, where the sourcees of the uncertainties tagged with $'{\ast}'$ are assumed to be 100\% correlated among the three different $D^0$ decay modes.
The efficiencies for other selection criteria, the trigger simulation, the event start time determination and the FSR simulation are quite high ($>99\%$), their systematic uncertainties are estimated to be less than 1\% ~\cite{otherserrors}.
The total systematic uncertainties are given by the quadratic sum of the individual uncertainties, assuming all sources to be independent.
The uncertainty due to the fit procedure is considered during the upper limit determination described in the following.

 \begin{table*}[htbp]
   \begin{center}
  \caption{Summary of systematic uncertainties (in \%) for $\jpsi \to D^{0}\epem$ and $\psi \to D^{0}\epem$, the sources tagged with $'{\ast}'$ are correlated among the different $\D0 $ decay modes. The hyphen ($-$) indicates the source does not contribute to the channel.}\label{tab:uncertaintypsi}
  \small
  \begin{tabular}{ l|c|c|c|c|c|c}
  \hline \hline
  \multirow{2}{*}{}&
            \multicolumn{2}{ c}{~~$D^{0}\to K^{-}\pi^{+}$}~~&
            \multicolumn{2}{|c}{~~$D^{0}\to K^{-}\pi^{+}\piz$}~~&
            \multicolumn{2}{|c}{~~$D^{0}\to K^{-}\pi^{+}\pppm$}~~\\
   \cline{2-7}
            & ~~~~~$\jpsi$~~~~~ & ~~~$\psip$~~~ & ~~~~~$\jpsi$~~~~~ & ~~~$\psip$~~~ &~~~~~$\jpsi$~~~~~ & ~~$\psip$~~ \\
   \hline
   Tracking$^{\ast}$             & 4.0 &4.0 & 4.0& 4.0 &6.0 & 6.0  \\
   PID$^{\ast}$               & 6.0 & 6.0 & 6.0& 6.0 &8.0&8.0   \\
   $\gamma$  detection         & --&--&1.2 & 1.2 & -- &--        \\
   Kinematic fit               &1.7 &1.6 &1.1 & 1.8&2.2 &2.0     \\
   Veto $\gamma$ conversion$^{\ast}$&1.7 & 1.7&1.7&1.7 &1.7&1.7 \\
   Veto $K_S\to\piz \piz$       &--  &-- &0.6 &--&--&             \\
   Veto $K_S\to\pi^{+}\pi^{-}$ &-- & -- &-- & -- &2.1&2.2     \\
   Veto $\jpsi\to\epem$        &-- &0.1 &--&--& &--              \\
   Branching fraction          &1.3 & 1.3&3.6 &3.6 &2.6&2.6      \\
   $\psi$ total number$^{*}$ &0.55&0.62 &0.55& 0.62&0.55&0.62    \\
   Others & 1.0 & 1.0& 1.0 &1.0 & 1.0 &1.0\\  \hline
   Total & 7.8 &7.8 & 8.5&8.7  & 11.0&10.9\\  \hline \hline
  \end{tabular}
  \end{center}
  \end{table*}

Since no significant signal for $\psi \to D^{0}\epem$ is observed, upper limits at the 90\% C.L.\ on the branching fractions are determined.
Simultaneous, unbinned maximum likelihood fits on the distributions of invariant masses
$M(\kpi)$, $M(\kpipi)$, and $M(\kpipipi)$, are carried out for the $\jpsi$ and $\psip$ samples.
In the fit, the signal shapes are described by the corresponding signal MC samples and the background shapes are described by 2nd order polynomial functions.
The expected number of signal events in the $i^{\rm th}$ decay mode is calculated with $N_{i} = N_{\psi}~\cdot~\Br~\cdot~\Br^{\rm inter}_{i}~\cdot~ \epsilon_i$,
where $N_{\psi}$ is the total number of $\psi$ events, $\Br_{i}^{\rm inter}$ is the product of the decay branching fractions of $\D0 $ mesons and subsequent intermediate states, taken from the PDG~\cite{PDG}, and $\epsilon_i$ is the detection efficiency from the signal MC samples.  The decay branching fraction $\Br$ of $\psi\to\D0 \epem$ is a common parameter among the three $\D0 $ decay modes.
The overall likelihood values ($\mathcal{L}$) are the products of those of the three $\D0 $ decay modes, incorporating systematic uncertainties, which are separated as correlated and uncorrelated~\cite{upper1,upper2}.
The likelihood fits are carried out with the MINUIT package~\cite{minuit}.

We compute the upper limits on the branching fraction at the 90\% C.L.\ using a Bayesian method~\cite{PDG} with a flat prior.
The optimized likelihoods, $\mathcal{L}$, are presented as a function of branching fraction $\Br(\psi\to\D0 \epem)$.
The upper limits on the branching fractions $\Br^{\rm UP}$ at the 90\% C.L.\ are the values that yield 90\% of the likelihood integral over $\Br$ from zero to infinity:  $\int_0^{\Br^{\rm UP}}{\mathcal{L}d\Br}/\int_0^\infty{\mathcal{L}d\Br}=0.9$.
To take into account the systematic uncertainties related to the fit process, two alternative fit scenarios are considered:
(1) changing the fit range on the invariant masses by 10 $\mevcc$; or
(2) replacing the 2nd order polynomial function with a 3rd order polynomial function for the background.
We try all combinations of the different scenarios. The one with the maximum upper limits on the branching fractions is taken as the conservative result.
The upper limits at the 90\% C.L.\ on the branching fractions are $\Br(\jpsi\to\D0 \epem) < 8.5\times 10^{-8}$ and $\Br(\psip\to\D0 \epem)< 1.4\times 10^{-7}$, respectively.
The corresponding normalized likelihood distributions are shown in Fig.~\ref{reftotalupperjpsi} and the best fit curves are shown in Fig.~\ref{masspsi}.

\begin{figure}[htb]

\setlength{\abovecaptionskip}{0.2cm}
\setlength{\belowcaptionskip}{-0.5cm}
\vskip  0.2cm
  \begin{overpic}[width=0.23\textwidth, height=0.20\textwidth]{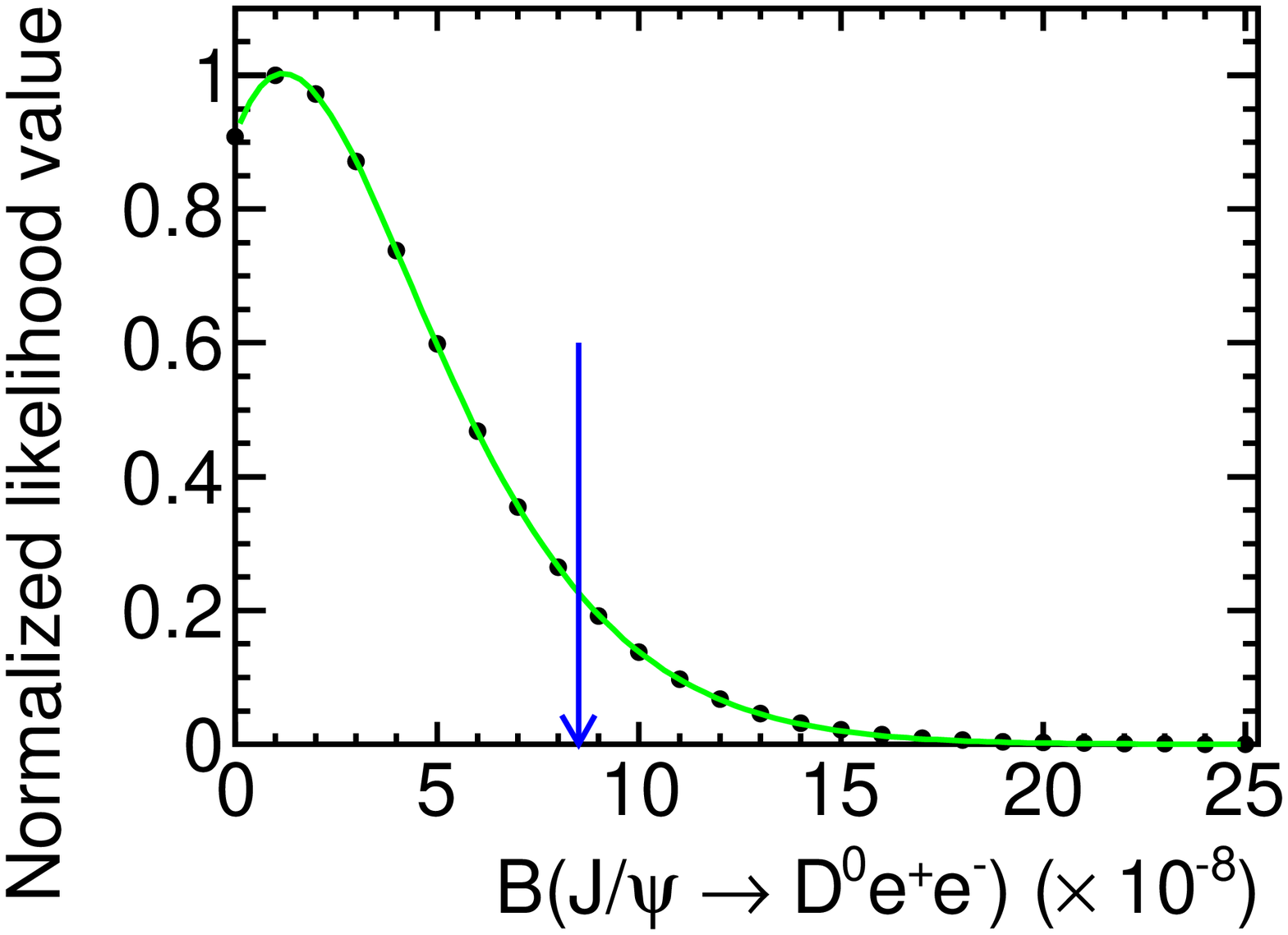}
  \end{overpic}
  \begin{overpic}[width=0.23\textwidth, height=0.20\textwidth]{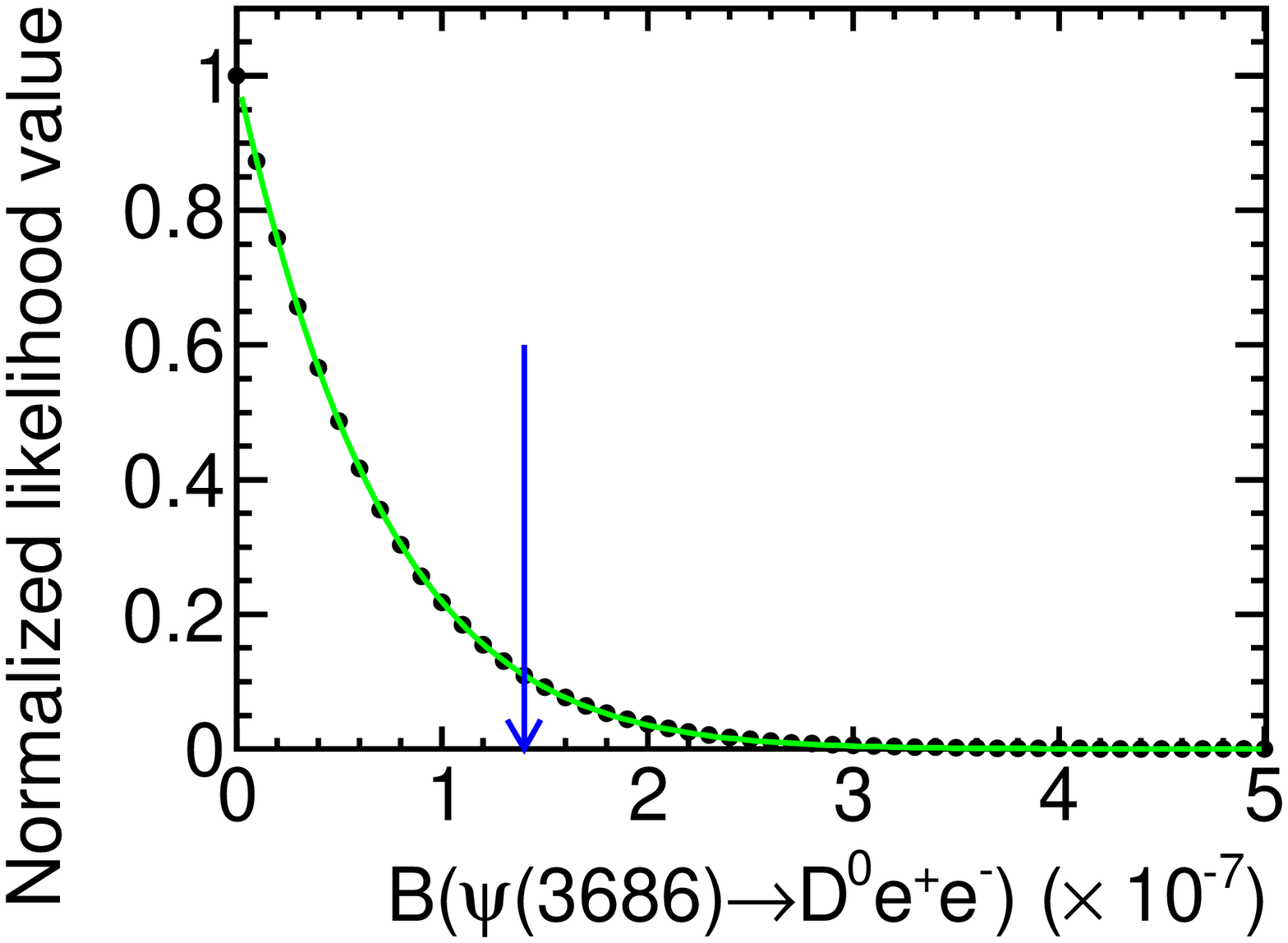}
  \end{overpic}
  \caption{Normalized combined  likelihood as a function of the decay branching fraction $\Br(\psi\to \D0 \epem)$ for $\jpsi$ (left) and $\psip$ (right) samples, where the correlated and un-correlated systematic uncertainties are incorporated. The likelihood function is normalized with the maximum to be 1. The blue arrow denotes the 90\% C.L.}\label{reftotalupperjpsi}

\end{figure}

In summary, we perform a search for the rare decays of $\jpsi \to \D0 \epem$ and $\psip\to \D0 \epem$
using samples of ($1310.6 \pm 7.2 $) $\times$ $10^{6}$ $J/\psi$ events and ($448.1 \pm 2.9$) $\times$ $10^{6}$ $\psip$ events collected with the BESIII detector.
No significant signal is observed and upper limits at the 90\% C.L.\ for the branching fractions are determined to be $\Br(\jpsi \to \D0 \epem)< 8.5\times 10^{-8}$ and $\Br(\psip \to \D0 \epem)<1.4\times 10^{-7}$, respectively.
The limit on $\Br(\jpsi\to\D0 \epem)$ is more stringent by two orders in magnitude compared to
the previous results, and the $\Br(\psip\to\D0 \epem)$ is set for the first time.
Though the upper limits are larger than the SM predictions, they may help to discriminate the different new physics models or to constrain the parameters in the different physics models.
Additionally, higher statistics $\jpsi$ and $\psip$ samples may help to improve the sensitivity of the measurements.

The BESIII collaboration thanks the staff of BEPCII, the IHEP computing center and the Supercomputing center of USTC for their strong support. This work is supported in part by National Key Basic Research Program of China under Contract No. 2015CB856700; National Natural Science Foundation of China (NSFC) under Contracts Nos. 11235011, 11322544, 11335008, 11425524, 11322544, 11375170, 11275189, 11475169, 11475164, 11575133, 11625523, 11635010; the Chinese Academy of Sciences (CAS) Large-Scale Scientific Facility Program; the CAS Center for Excellence in Particle Physics (CCEPP); the Collaborative Innovation Center for Particles and Interactions (CICPI); Joint Large-Scale Scientific Facility Funds of the NSFC and CAS under Contracts Nos. 11179007, U1332201, U1532257, U1532258; CAS under Contracts Nos. KJCX2-YW-N29, KJCX2-YW-N45, QYZDJ-SSW-SLH003; 100 Talents Program of CAS; National 1000 Talents Program of China; INPAC and Shanghai Key Laboratory for Particle Physics and Cosmology; German Research Foundation DFG under Contracts Nos. Collaborative Research Center CRC 1044, FOR 2359; Istituto Nazionale di Fisica Nucleare, Italy; Koninklijke Nederlandse Akademie van Wetenschappen (KNAW) under Contract No. 530-4CDP03; Ministry of Development of Turkey under Contract No. DPT2006K-120470; National Natural Science Foundation of China (NSFC) under Contracts Nos. 11505034, 11575077; National Science and Technology fund; The Swedish Research Council; U. S. Department of Energy under Contracts Nos. DE-FG02-05ER41374, DE-SC-0010118, DE-SC-0010504, DE-SC-0012069; University of Groningen (RuG) and the Helmholtzzentrum fuer Schwerionenforschung GmbH (GSI), Darmstadt; WCU Program of National Research Foundation of Korea under Contract No. R32-2008-000-10155-0



\begin{thebibliography}{99}


 \bibitem{GIM}S.~L.~Glashow, J.~Iliopoulos and L.~Maiani, Phys.\ Rev.\ D {\bf 2}, 1285 (1970).
 \bibitem{FCNC}Y.~M.~Wang {\it et al.},  J. Phys. G {\bf 36}, 105002  (2009).
 \bibitem{SM1}M.~A.~Sanchis-Lonzano, Z. Phys.\ C {\bf 62}, 271 (1994).
 \bibitem{SM2}Y.~M.~Wang {\it et al.}, Eur.\ Phys.\ J.\ C {\bf 54}, 107 (2008) .
 \bibitem{TopC}C.~Hill, Phys.\ Lett.\ B {\bf 345}, 483 (1995).
 \bibitem{Supersy}C.~S.~Aulakh and R.~N.~Mohapatra, Phys.\ Lett.\ B {\bf 119}, 136 (1982).
 \bibitem{Thiggs}S.~Glashow and S.~Weinberg, Phys.\ Rev.\ D {\bf 15}, 1858 (1977).
 \bibitem{newphy1} X.~Zhang, HEP\&NP, {\bf 25}, 461 (2001).
 \bibitem{newphy2}A.~Datta, P.~J.~Odonnel, S.~Pakvasa and X.~Zhang, Phys.\ Rev.\ D {\bf 60}, 014011 (1999).
 \bibitem{besII} M.~Ablikim {\it et al.} [BES Collaboration], Phys.\ Lett.\ B {\bf 639},  418 (2006).
 \bibitem{detector1} M.~Ablikim {\it et al.} [BESIII Collaboration], Nucl.\ Instrum.\ Meth.\ A {\bf 614}, 3 (2010).
 \bibitem{jpsinum1}  M.~Ablikim {\it et al.} [BESIII Collaboration], Chin.\ Phys.\ C {\bf 36}, 915 (2012).
 \bibitem{jpsinum2}  M.~Ablikim {\it et al.} [BESIII Collaboration], Chin.\ Phys.\ C {\bf 41}, 013001 (2017).
 \bibitem{psipnum1} M.~Ablikim {\it et al.} [BESIII Collaboration], Chin.\ Phys.\ C  {\bf 37}, 063001 (2013).
 \bibitem{psipnum2} M.~Ablikim {\it et al.} [BESIII Collaboration],
   arXiv:1709.03653 [hep-ex], submitted to Chin.\ Phys.\ C.
 \bibitem{geant4} S.~Agostinelli {\it et al.} [{\sc geant4} Collaboration], Nucl \ Instrum.\ Meth.\ A {\bf 506}, 250 (2003).
 \bibitem{Deng} Z.~Y.~Deng {\it et al.},  Chin.\ Phys.\ C {\bf 30}, 371 (2006).
 \bibitem{kkmc1} S.~Jadach, B.~F.~L.~Ward and Z.~Was, Comput. Phys. Commun. {\bf 130}, 260 (2000); Phys.\ Rev.\ D {\bf 63}, 113009  (2001).
\bibitem{evtgen} D. J.~Lange, Nucl.\ Instrum.\ Meth. A {\bf462}, 152 (2001); R.~G.~Ping {\it et al.}, Chin.\ Phys.\ C {\bf 32}, 599 (2008).
\bibitem{PDG} C.~Patrignani {\it et al.} [Particle Data Group], Chin.\ Phys.\ C, {\bf 40}, 100001  (2016).
\bibitem{lundcharm} J.~C.~Chen {\it et al.}, Phys.\ Rev.\ D {\bf 62}, 034003 (2000).

\bibitem{photos} E.~Barberio and Z.~Was, Comput.\ Phys.\ Commun.\ {\bf 79}, 219 (1994).
\bibitem{DIYg} V.~M.~Budnev and V.~A.~Karnakov, Pisma Zh. Eksp. Teor. Fiz. {\bf29}, 439 (1979).
\bibitem{xuzr} Z.~R.~Xu and K.~L.~He, Chin.\ Phys.\ C {\bf 36}, 742 (2012).

\bibitem{trackerror1} M.~Ablikim {\it et al.} [BESIII Collaboration], Phys.\ Rev.\ D {\bf 83}, 112005 (2011).
\bibitem{trackerror2} M.~Ablikim {\it et al.} [BESIII Collaboration], Phys.\ Rev.\ D {\bf 85}, 092012 (2012).

\bibitem{kinematic} M.~Ablikim {\it et al.} [BESIII Collaboration], Phys.\ Rev.\ D {\bf 87}, 012002 (2013).
\bibitem{rhopi} R.~G.~Ping, Gang Li and Z.~Wang, Formalism of Helicity Coupling Amplitudes for $\jpsi \to \pi^{+} \pi^{-} \pi^{0}$. Commun.\ Theor.\ Phys.\ (Beijing, China) {\bf47} (2007).
\bibitem{otherserrors} M.~Ablikim {\it et al.} [BESIII Collaboration], Phys.\ Rev.\ Lett. {\bf110}, 252002 (2013).
\bibitem{upper1} M.~R.~Convery, Incorporating multiplicative systematic errors in branching ratio limits. SLAC-TN-03-001, 2003.
\bibitem{upper2} K.~Stenson, A more exact solution for incorporating multiplicative systematic uncertainties in branching ratio limits. arXiv:0605236 (2006).
\bibitem{minuit} F.~James and M.~Roos, Comput.\ Phys.\ Commun. {\bf 10}, 343 (1975).
\end{thebibliography}
\end{document}